\documentclass[%
 reprint,
amsmath,amssymb,
aps,
pra,
]{revtex4-2}

\usepackage{graphicx}
\usepackage{dcolumn}
\usepackage{bm}
\usepackage{color}
\usepackage{upgreek}
\usepackage{CJK}
\usepackage[%
  colorlinks=true,
  urlcolor=blue,
  linkcolor=blue,
  citecolor=blue
]{hyperref}
\usepackage{siunitx}

\DeclareSIUnit\electrons{e\textsuperscript{--}}
\DeclareSIUnit\neutrons{neutrons}
\DeclareSIUnit\ppm{ppm}
\DeclareSIUnit\ppb{ppb}
\DeclareSIUnit\lines{l}
\DeclareSIUnit{\calorie}{cal}

\begin{document}

\begin{CJK*}{UTF8}{gbsn}

\preprint{APS/123-QED}

\title{Delayed luminescence and thermoluminescence in laboratory-grown diamonds}

\author{Jiahui Zhao (赵嘉慧)}
 \email{gloria.zhao.1@warwick.ac.uk}
 \affiliation{Department of Physics, University of Warwick, Coventry CV4 7AL, United Kingdom
}
\author{Ben G. Breeze}
  \affiliation{Spectroscopy Research Technology Platform, University of Warwick, Coventry CV4 7AL, United Kingdom}
\author{Ben L. Green}
\affiliation{Department of Physics, University of Warwick, Coventry CV4 7AL, United Kingdom
}

\author{Hengxin Yuan (原亨馨)}
\affiliation{Zhengzhou Sino-Crystal Diamond Co., Ltd., Zhengzhou 450001, China}

\author{Troy Ardon}
\affiliation{Gemological Institute of America (GIA), New York NY10036, USA}

\author{Wuyi Wang (王五一)}
\affiliation{Gemological Institute of America (GIA), New York NY10036, USA}

\author{Mark E. Newton}
 \email{m.e.newton@warwick.ac.uk}
\affiliation{Department of Physics, University of Warwick, Coventry CV4 7AL, United Kingdom
}
\maketitle
\end{CJK*}




\date{\today}

\begin{abstract}
The ``blue-green" phosphorescence/thermoluminescence is most commonly observed in diamonds following excitation at or above the indirect band gap and has been explained by a substitutional nitrogen-boron donor-acceptor pair recombination model \cite{zhao2023}. ``Orange" and ``red" phosphorescence have also been frequently observed in lab-grown near-colourless high-pressure high-temperature diamonds following optical excitation, and their luminescence mechanisms are shown to be different from that of the ``blue-green" phosphorescence. The physics of the ``orange" and ``red" luminescence and phosphorescence bands including the optical-excitation dependency (UV-NIR), temperature dependency (20 - 573 K), and related charge transfer process are investigated by a combination of self-built time-resolved imaging/spectroscopic techniques. In this paper, an alternative model for long-lived phosphorescence based on charge trapping is proposed to explain the ``orange" phosphorescence/thermoluminescence band. Additionally, the ``red" phosphorescence band are attributed to point defect which possibly has a three-level phosphorescence system.
\end{abstract}

\maketitle


\section{Introduction}

The ``blue-green", ``orange", and ``red" phosphorescence in diamonds presenting broad featureless spectroscopic bands following optical or X-ray excitation at room temperature are commonly reported in the literature \cite{eaton2008,eaton2011,watanabe1997phosphorescence,ulrika2015,krumme1964,walsh1971,cheng2023synthetic}. The ``blue-green" emission band, with a peak energy of $\sim$ 2.5~eV (490 nm - 503 nm), is commonly observed in type II natural, high-pressure high-temperature (HPHT) synthetic, and HPHT-treated diamonds \cite{eaton2008,eaton2011,krumme1964,chandrasekharan1946,chandrasekharan1946patterns,walsh1971,song2016,watanabe1997phosphorescence,ulrika2015,shao2020}, and originates from neutral substitutional nitrogen-boron donor-acceptor pairs ($\text{N}_\text{S}^0$...$\text{B}_\text{S}^0$) recombination \cite{zhao2023}.

The peak energy of the ``orange" or ``yellowish-orange" phosphorescence band ranges from 575 nm \cite{eaton2008,ulrika2015} to 590 nm \cite{watanabe1997phosphorescence,walsh1971}, is often referred to as the 2.1 eV band. This ``orange" emission band can be observed in HPHT synthetic or HPHT treated diamonds after UV or visible excitation \cite{eaton2008,eaton2011,ulrika2015,watanabe1997phosphorescence,walsh1971}. An ``orange" phosphorescence band centred at 580 nm was reported to have been seen in natural diamond \cite{ulrika2015}. Compared to the ``blue-green" band centred at $\sim$ 2.5 eV, the ``orange" phosphorescence band is not common, has a weaker initial intensity and persists for several seconds to tens of seconds \cite{ulrika2015,watanabe1997phosphorescence}. In samples containing this emission band, after excitation the phosphorescence can be seen to change from blue to orange. The ``orange" emission band and ``blue-green" emission band are reported to have different energy excitation thresholds, so are likely to have different origins \cite{watanabe1997phosphorescence}.

Like the ``orange" emission band, a ``red" phosphorescence band has peak energies ranging from 1.8 - 1.9~eV (660 - 690 nm) \cite{eaton2008,eaton2011,krumme1964,walsh1971}, observed in natural diamonds is always accompanied by the ``blue-green" band centred at $\sim$ 2.5 eV. The ``red" phosphorescence can be excited by either UV or visible light \cite{eaton2008}. The decay of the ``red" emission band has a lifetime ranging from several seconds to tens of seconds, and is slower than that of the 2.5 eV ``blue-green" band \cite{eaton2008}. As the concentration of boron in diamond increases, the ratio of the initial intensity of the ``blue-green" band and the ``red" band decreases, and the half-lifetime of phosphorescence bands becomes shorter \cite{eaton2011}. Sometimes the ``red" phosphorescence cannot be seen until the diamond is heated up to 350 K \cite{krumme1964}. A weak ``red" phosphorescence band was seen in an HPHT synthetic diamond grown from a cobalt solvent using titanium as the nitrogen ``getter" by Watanabe \emph{et al.} \cite{watanabe1997phosphorescence}. 

Phosphorescence and thermoluminescence share similar mechanisms \cite{chen2003,mckeever1988thermoluminescence} but discussion of the correlation between phosphorescence and thermoluminescence in diamond, and the comparison of optical activation energy and thermal activation energy is rare in the literature.

Thermoluminescence studies on diamonds where the samples were illuminated at liquid nitrogen temperature (approximately 80 K) are summarised (see supplemental information) \cite{chandrasekharan1946TL,bull1950,halperin1961,halperin1966,walsh1971,levinson1973,bourgoin1978,petitfils2007,zhao2023}. ``Blue-green" and ``red" thermoluminescence is seen in natural, HPHT synthetic and boron-doped CVD synthetic diamonds. Since the published TL spectra are limited, it is uncertain whether the same colour TL emission (``blue-green" or ``red") in different samples originates from the same defect. The ``red" thermoluminescence band peak at 1.85 eV reported by Walsh \emph{et al.} \cite{walsh1971} is very similar to the ``red" phosphorescence band \cite{eaton2008,eaton2011}. The ``blue-green" TL band has been linked to the ``blue-green" phosphorescence band \cite{zhao2022,shao2020}. The same defects may play roles in both thermoluminescence and phosphorescence \cite{walsh1971}. 

``Red" and ``blue-green" thermoluminescence appear in the temperature range belonging to different TL glow peaks, corresponding to traps with different activation energies \cite{nahum1963,halperin1966,walsh1971,levinson1973,bourgoin1978}. This indicates that the carriers thermally released from one type of trap can give rise to emission from different luminescence centres \cite{halperin1966}. The emission from the same type of luminescence centre can also be generated by carriers from different traps.

Two TL glow peaks corresponding to traps with activation energies of approximately 0.2~eV and 0.37 eV are commonly observed in type II natural and HPHT synthetic diamonds \cite{nahum1963,halperin1966,walsh1971,levinson1973,bourgoin1978}. The thermal activation energy is usually calculated by various methods (see \cite{zhao2022,mckeever1988thermoluminescence}), such as the initial rise method. However, in cases where TL peaks overlap, or the thermal scan of the TL glow peak does not start with zero intensity, the initial rise region is not well-defined, and this method does not result in the determination of accurate activation energies (e.g. \cite{nahum1963}). Thermal cleaning can be applied to mitigate this problem. The relative intensity of these two glow peaks (0.2 eV and 0.37 eV) is sample dependent, indicating they correspond to different traps in diamonds \cite{halperin1966}. These two traps have been assigned to negatively charged substitutional nitrogen donor $\text{N}_\text{S}^-$ ($\sim$ 0.2~eV) and boron acceptor $\text{B}_\text{S}^0$ (0.37 eV), respectively \cite{zhao2023}.

\section{Experimental details}

\subsection{Samples}

Three near colourless HPHT grown diamonds are the focus of this paper, two of which, Sino-01 and SYN4-10, were supplied as brilliant cut gems. Samples GE81-107a-B and Sino-01 were grown from iron-cobalt solvent-catalyst systems, but the composition of the solvent-catalyst for SYN4-10 is not known. The nature of the nitrogen ``getters" used is not known, but the most abundant impurities in all samples were single substitutional nitrogen and boron. The concentration of neutral substitutional nitrogen ([$\text{N}_\text{S}^0$]) was measured by Electron Paramagnetic Resonance (EPR) and the concentration of neutral substitutional boron ([$\text{B}_\text{S}^0$]) by Fourier Transform infrared (FTIR) absorption (see Table \textcolor{blue}{\ref{Table:samples concentrations}}). In all samples the concentration of $\text{N}_\text{S}^0$ and $\text{B}_\text{S}^0$ (considering all charge states) varied between 10's and 100's of ppb, but both impurities were heterogeneously distributed (significant variations between different growth sectors) making quantitative analysis difficult. Both Sino-01 and SYN4-10 samples exhibit photoluminescence (PL) and/or cathodoluminescence (CL) emission features attributed to nickel defects (e.g. Sino-01: CL zero phonon line (ZPL) at 484 nm and 703.6 nm \cite{zaitsev2013}; SYN4-10: PL at 484 nm, 679.5 nm and doublet at 884 nm \cite{yelisseyev2003, dobrinets2016}) indicating that the solvent-catalyst contained at least trace amounts of nickel. No nickel related defects were detected in GE81-107a-B sample by PL or CL.

\begin{table}
\renewcommand\arraystretch{1.3}
\centering
\caption{The concentrations of $\text{B}_\text{S}^0$ and $\text{N}_\text{S}^0$ in samples GE81-107a-B and SYN4-10 in the ``meta-stable ambient state".}
\begin{ruledtabular}
\begin{tabular}{ccc}
Concentration (ppb) & $[\text{B}_\text{S}^0]$ by FTIR & $[\text{N}_\text{S}^0]$ by EPR \\ \hline
GE81-107a-B & 67 $\pm$ 10 & 145 $\pm$ 20 \\
SYN4-10 & 350 $\pm$ 20 & 15.5 $\pm$ 5 \\
\end{tabular}
\label{Table:samples concentrations}
\end{ruledtabular}
\end{table}

\subsection{Experimental setup}

\subsubsection{Variable temperature photoluminescence}

Photoluminescence spectra ranging from 250 - 800 nm excited by optical excitation at wavelengths between 200 - 450 nm were collected using an Edinburgh Instrument FS5 spectrometer equipped with a photon counting photo-multiplier. Phosphorescence spectra was recorded at 20 K using an optistat (CF,  Oxford Instruments) with a temperature controller (ITC503, Oxford Instruments).

\subsubsection{Phosphorescence and thermoluminescence}

Phosphorescence and thermoluminescence emission were investigated by a variable-temperature time-gated luminescence experimental system referred to as ``Garfield" previously described \cite{zhao2022}. A pulsed 224 nm laser (HEAG70-224SL, Photon Systems) with 50 mW peak power, or a 375 nm continuous wave (CW) laser (L375P70MLD, Thorlabs) with 70 mW power, were used as the optical excitation source. A variable temperature stage (THMS600 Linkam) enabled studies at temperatures between 83 K to 673 K. The luminescence was detected by either a fibre-coupled spectrometer (ANDOR Shamrock i303) operating between 400 - 1000 nm, or a camera (CMOS, HAMAMATSU, C11440-36U) operating between 300 - 1100 nm. The hardware components, including the laser, temperature stage, and detectors, are controlled by an Arduino platform which can achieve $\sim$ms timing accuracy. 

Phosphorescence spectra were collected following the optical excitation with the 224 nm pulsed laser described earlier to ``saturation" over a temperature range of 83 - 573 K at intervals of 20 K. For the brilliant-cut samples Sino-10 and SYN4-10, the phosphorescence displayed clear growth sector dependence, making it impossible to calculate activation energies using any phosphorescence decay curve fitting methods. 

To study the thermoluminescence, diamond samples were firstly optically excited to saturation at 83 K and then heated to 673 K at a linear rate of 100 K/min after the phosphorescence decayed to zero. TL was recorded by the camera or spectrometer. The ``thermal cleaning" method was used to deconvolute overlapping TL glow peaks and the activation energies of each corresponding TL peak were estimated by the initial rise method \cite{mckeever1988thermoluminescence,zhao2022}.

\subsubsection{UV-excited EPR}

Rapid-passage EPR (Bruker E580 spectrometer coupled to an X-band microwave bridge) was used to measure the average concentration of $\text{N}_\text{S}^0$ before but with different light source \cite{zhao2023}, in this study the measurement was performed during and after above bandgap optical excitation by a 60 W xenon flash lamp (L7685, Hamamatsu) with 228 nm bandpass filter (FWHM 25 nm) in diamond sample SYN4-10.

\section{Results}

\subsection{Photoluminescence}

\begin{figure}[htbp]
    \centering
    \includegraphics[width=\linewidth]{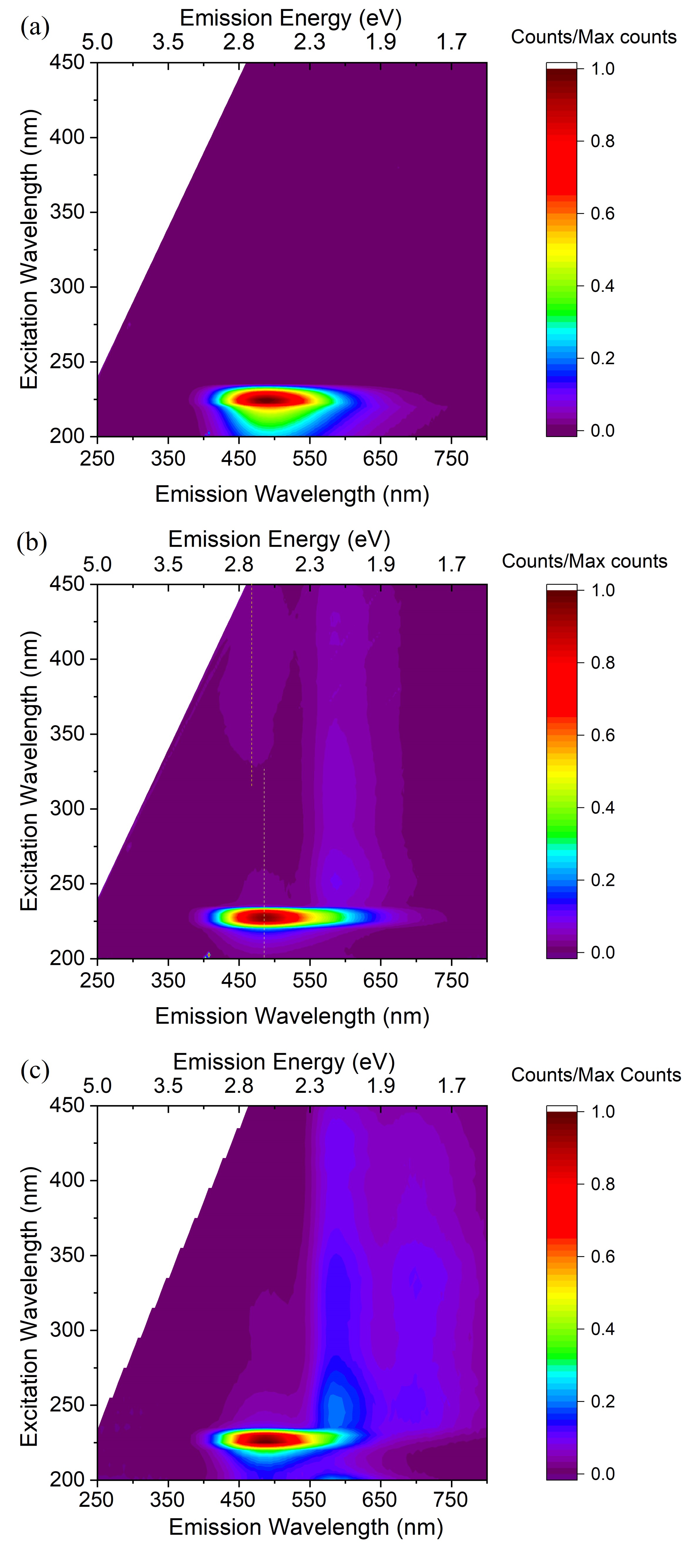}
    \caption{Room temperature photoluminescence excitation plots recorded with FS5 instrument (a) GE81-107a-B (b) Sino-01 (c) SYN4-10.}
    \label{fig:Em vs Ex}
\end{figure}

At room temperature all samples exhibited an intense broad ``blue-green" luminescence band upon optical excitation of wavelengths less than 236 nm, as shown in Fig \textcolor{blue}{\ref{fig:Em vs Ex}}. This band originates from donor-acceptor pair recombination from $\text{N}_\text{S}^0$...$\text{B}_\text{S}^0$ pairs \cite{zhao2023}. In both Sino-01 and SYN 4-10 samples at room temperature, an additional broad ``orange" emission band centred at $\sim$ 2.1 eV is observed with excitation ranging from 200-450 nm (Fig \textcolor{blue}{\ref{fig:Em vs Ex}}). In samples SYN4-10 at room temperature an additional broad ``red" emission band centred at $\sim$ 1.8 eV is observed with excitation wavelengths longer than 220 nm.

\subsection{Phosphorescence}

\subsubsection{Sample GE81-107a-B}

The ``blue-green" phosphorescence from sample GE81-107a-B has been previously described \cite{zhao2023}. This band is excited with light of wavelengths $<$ 236 nm and originates from $\text{N}_\text{S}^0$...$\text{B}_\text{S}^0$ donor-acceptor pairs recombination. Upon removal of the excitation in this sample, this phosphorescence can be observed for $\sim$ 100 s at room temperature, but the lifetime can vary from a few milliseconds to minutes, depending on the concentration of  $\text{N}_\text{S}$ and $\text{B}_\text{S}$ present in the sample \cite{zhao2023}.

\subsubsection{Sample Sino-01}

\begin{figure}[htbp]
    \centering
    \includegraphics[width=\linewidth]{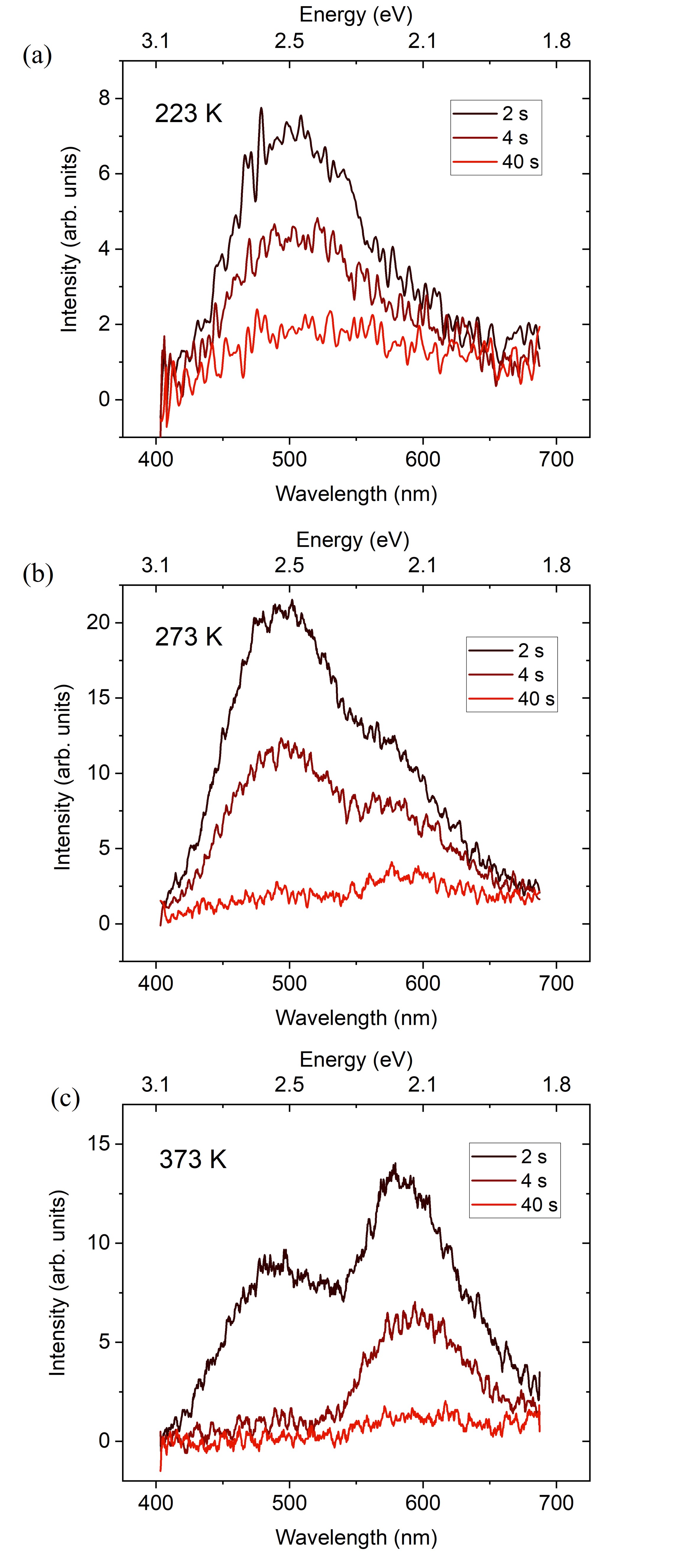}
    \caption[Phosphorescence spectra at different T]{Phosphorescence spectra after 224 nm excitation is measured by the ``Garfield" experimental setup with an integration time of 150 ms in sample Sino-01 at (a) 223 K, (b) 273 K, (c) 373 K.}
    \label{fig:224 Phos spectra sino-01}
\end{figure}

Two phosphorescence bands are observed after 224 nm excitation (Fig ~\textcolor{blue}{\ref{fig:224 Phos spectra sino-01}}): a ``blue-green" band centred at approximately 2.5 eV (493 nm) and an ``orange" band centred at 2.1 eV (580 nm - 590 nm). Similar ``blue-green" and ``orange" phosphorescence bands were observed in HPHT lab-grown diamonds previously \cite{eaton2011,watanabe1997phosphorescence,walsh1971}. 

At 223 K, one broad phosphorescence band at 2.5 eV (493 nm) is observed, decaying relatively slowly. There is no shift in band position during phosphorescence decay, and the emission lasts for tens of seconds. Above 253 K, the ``orange" phosphorescence band and the 2.5 eV ``blue-green" band are both observed. The lifetime of ``orange" emission is longer than that of ``blue-green" emission, consistent with the observation of Watanabe \emph{et al.} \cite{watanabe1997phosphorescence} (Fig ~\textcolor{blue}{\ref{fig:224 Phos spectra sino-01}(b)(c)}). The higher the temperature, the initial intensity of the ``orange" phosphorescence is stronger than that of the ``blue-green" band. The ``orange" band shifts slightly to lower energy as the temperature is increased. The ``orange" band shifts to lower energy (by approximately 0.04 eV) during phosphorescence decay. 

\subsubsection{Sample SYN4-10}

\paragraph{224 nm excitation}

\begin{figure}[htbp]
    \centering
    \includegraphics[width=\linewidth]{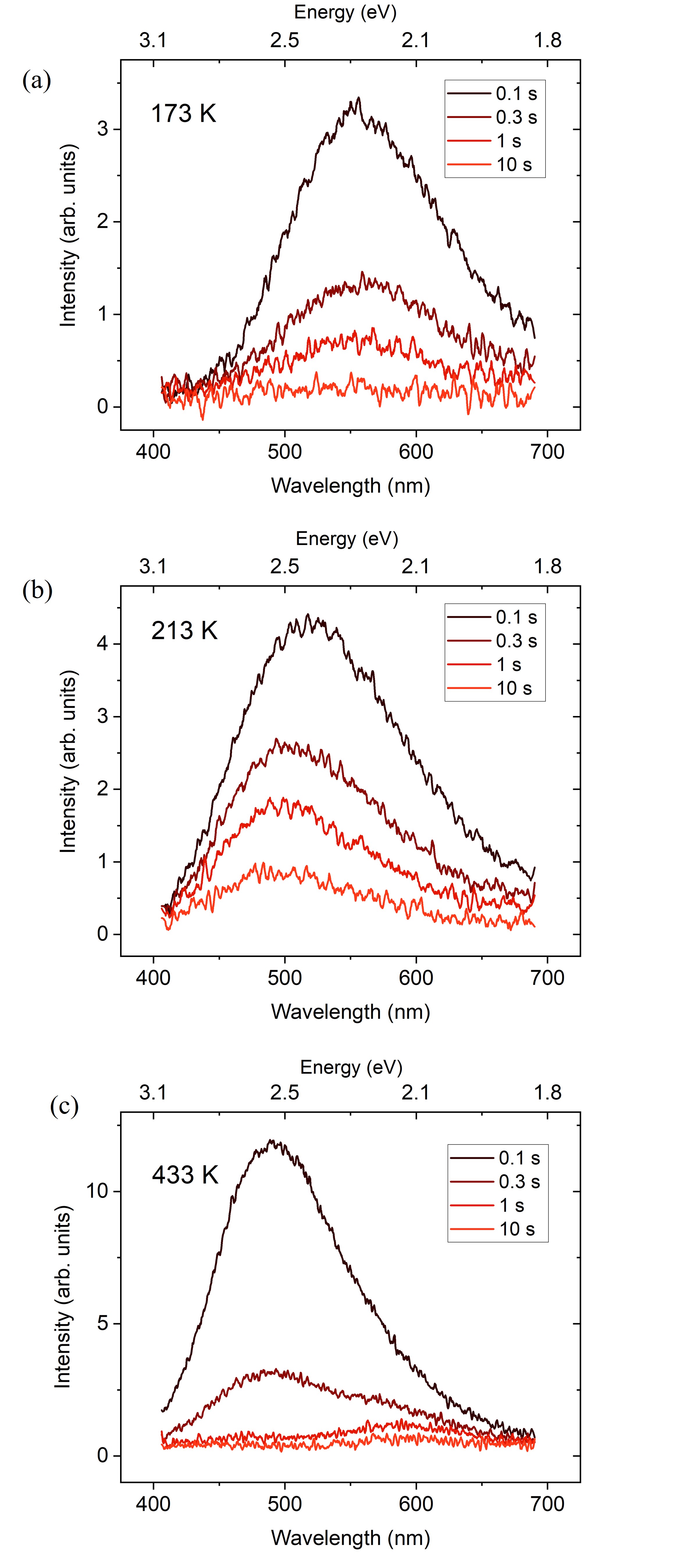}
    \caption{Phosphorescence spectra after 224 nm excitation is measured by the ``Garfield" experimental setup with an integration time of 150 ms in sample SYN4-10 at (a) 173 K, (b) 213 K, (c) 433 K.}
    \label{fig:224 Phos spectra syn4-10}
\end{figure}

The common ``blue-green" phosphorescence band was observed (Fig \textcolor{blue}{\ref{fig:224 Phos spectra syn4-10}}). In the temperature range of 83 - 173 K, the band is centred at 2.25 eV (550 nm) and did not shift during the phosphorescence decay (Fig \textcolor{blue}{\ref{fig:224 Phos spectra syn4-10}(a)}). In the temperature range of 173 - 273 K, the peak shifts to higher energies during the phosphorescence decay and as the temperature is increased. The peak emission at long delay times and higher temperatures is $\sim$ 2.5 eV (Fig \textcolor{blue}{\ref{fig:224 Phos spectra syn4-10}(b)}). This phosphorescence behaviour is similar to that reported previously \cite{zhao2023}. 

In addition to the ``blue-green" band, an ``orange" band with a peak energy at $\sim$ 2.1 eV (590 nm) is observed at temperatures above 273 K. The decay of the 2.1 eV ``orange" phosphorescence band is significantly slower than that the 2.5 eV ``blue-green" phosphorescence band and lasts for $>$ 12 hours (Fig \textcolor{blue}{\ref{fig:224 Phos spectra syn4-10}(c)}). However, in this sample it is uncertain whether the band position of the ``orange" band shifts during decay because of the weak intensity. 

The peak emission of the ``blue-green" and ``orange" luminescence bands at room temperature are consistent with those reported in the literature \cite{watanabe1997phosphorescence,walsh1971}.

\paragraph{375 nm excitation}

\begin{figure}[htbp]
    \centering
    \includegraphics[width=\linewidth]{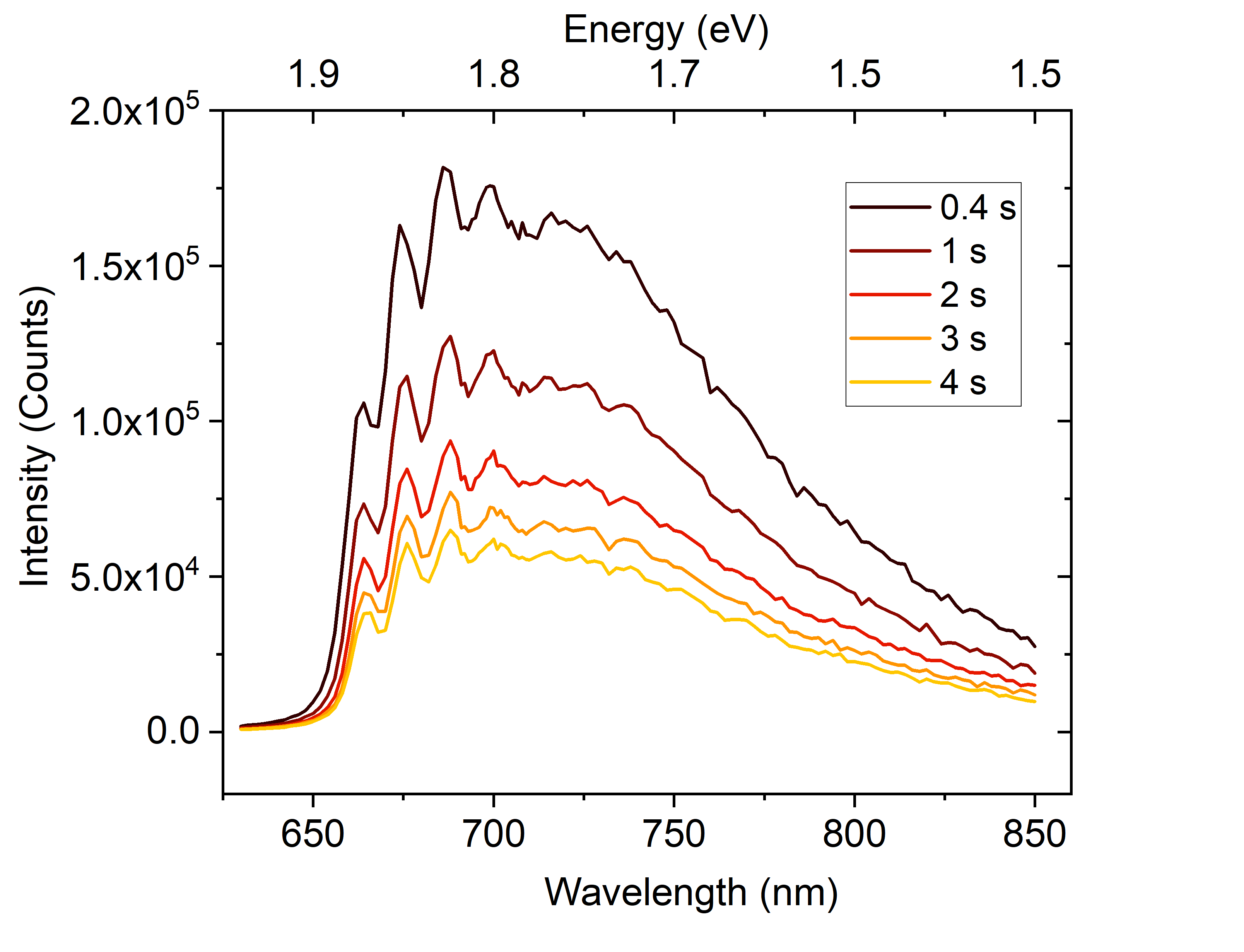}
    \caption{The phosphorescence spectra in sample SYN4-10 following 375 nm excitation recorded at 20 K.}
    \label{fig:SYN4-10 20K phos}
\end{figure}

\begin{figure}[htbp]
    \centering
    \includegraphics[width=\linewidth]{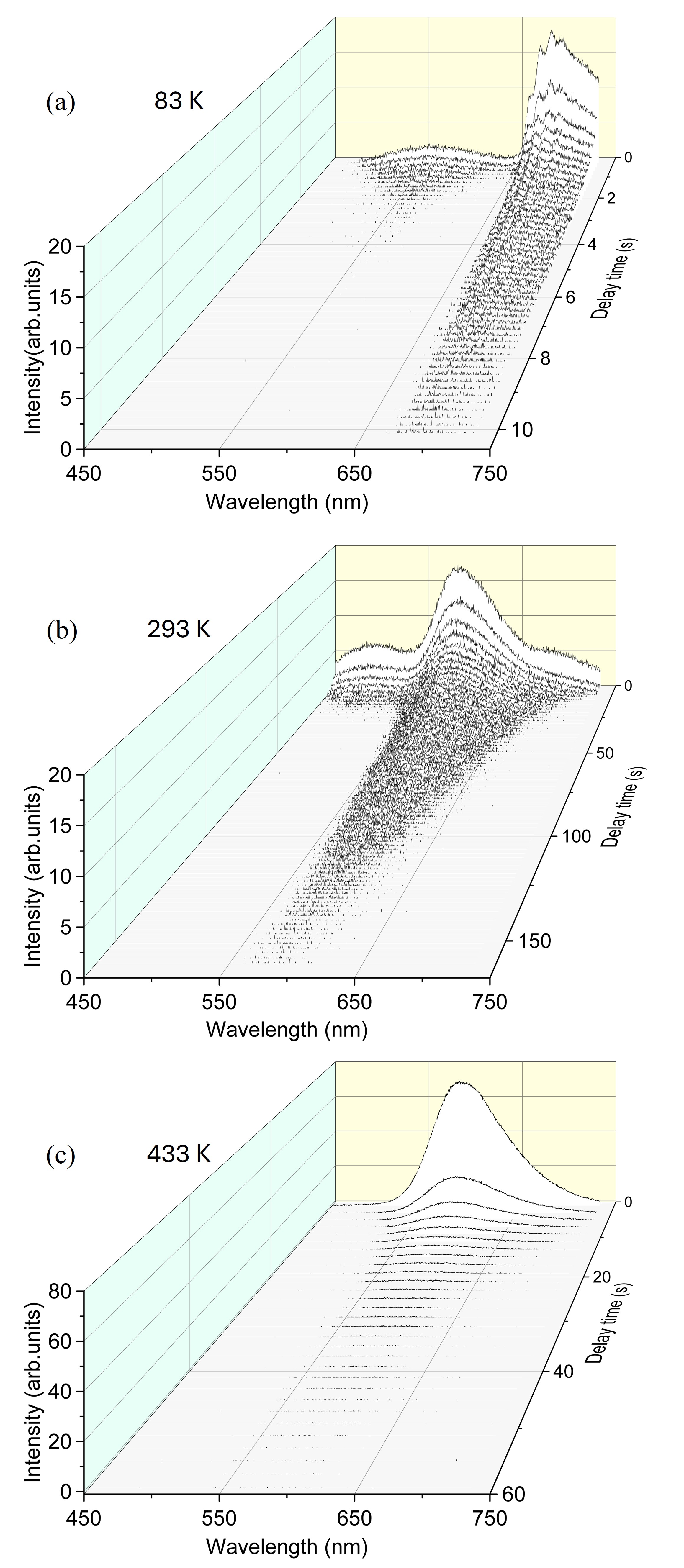}
    \caption{Phosphorescence spectra in sample SYN4-10 excited by 375 nm CW laser light recorded using Garfield experimental setup. (a) The 3D plot of phosphorescence decay at 83 K (b) 293 K (c)433 K.}
    \label{fig:SYN4-10 375 phos 3Dplots}
\end{figure}

Two phosphorescence bands are observed at 83 K as shown in Fig \textcolor{blue}{\ref{fig:SYN4-10 375 phos 3Dplots}(a)}: one weak band centred at 556 nm (2.23 eV) with a lifetime of $\sim$ 1 s and one much stronger ``red" phosphorescence band peak energies at 1.8 eV (680 nm) with a lifetime of around 10 s. The spectral position of this ``red" band is the same as Krumme's observation in a natural diamond \cite{krumme1964}. The ``red" phosphorescence spectra were recorded at 20~K (Fig \textcolor{red}{\ref{fig:SYN4-10 20K phos}}), the highest energy peak at 1.868 eV and vibronic replicas with a phonon energy of $\sim$ 31 meV are observed, more details will be discussed in section \textcolor{blue}{\ref{sec:red}}.

 Unlike other phosphorescence bands observed previously which are all broad and featureless, there is previously unreported resolved vibronic structure the ``red" emission band (Fig \textcolor{blue}{\ref{fig:SYN4-10 375 phos 3Dplots}(a)}) \cite{walsh1971,watanabe1997phosphorescence,ulrika2015,eaton2008,eaton2011,shao2020,su2018}. The vibronic structure of the ``red" band is not resolved at temperatures above 173 K. Meanwhile, the band position shifts to 690 nm, consistent with the ``red" phosphorescence band position report by Walsh \emph{et al.} \cite{walsh1971}. The centre position of the emission band (690 nm) does not change during the phosphorescence decay. The red-NIR phosphorescence band centred at 690 nm is observed over a temperature range of 173 K to 373 K; the initial intensity increases, and the lifetime decreases as temperature increases. When the temperature is above approximately 373 K, it is uncertain whether the ``red" band disappears or masked by the stronger ``orange" band.

 At least three emission bands can be simultaneously observed during the phosphorescence decay at temperatures between 253 K and 353 K (Fig \textcolor{blue}{\ref{fig:SYN4-10 375 phos 3Dplots}(c)}). All three bands show a significant shift to higher energies during the first $\sim$ 1~s of phosphorescence emission at RT. After that, the emission bands are centred at 490 nm (2.5 eV), 580 nm (2.1 eV), and 690 nm (1.8 eV), respectively. The 1.8 eV "red" band has the shortest phosphorescence lifetime, the 2.1 eV "orange" band the longest, and the lifetime of 2.5 eV band is in between. The peak of those three phosphorescence bands are observed at the same spectral position at higher temperatures up to around 373~K. The intensity of the ``orange" phosphorescence band increases dramatically with increasing temperature. 
 
 When the temperature is above 433 K (Fig \textcolor{blue}{\ref{fig:SYN4-10 375 phos 3Dplots}(d)}), the broad phosphorescence band centred at 590 nm is asymmetric. It has a long tail on the low energy side, which suggests that both the ``orange" and ``red" bands are contributing to the overall emission. The 2.5 eV ``blue-green" band is weak and not efficiently excited by the 375 nm excitation (note the 375 nm is 70 mW CW laser and the 224 nm pulsed laser has an average power of 0.05 mW).

\subsection{Thermoluminescence}                      

\subsubsection{Sample GE81-107a-B}

The thermoluminescence from sample GE81-107a-B has been described in details in our previous paper \cite{zhao2023}.It was shown for this sample that by delaying the start of the TL temperature ramp, defect centres which relax via phosphorescence can be eliminated from the subsequent TL glow curve. Accurate trap energies are difficult to determine for closely overlapping TL peaks, but the involvement of the boron acceptor ($\text{E}_\text{A}$ = 0.37 eV) and another trap with a lower energy was suggested from the analysis of TL data from different growth sectors of this sample, and confirmed by analysis of the recovery of the $[\text{N}_\text{S}^0]$ after the above bandgap optical excitation was removed \cite{zhao2023}.

\subsubsection{Sample Sino-01}

\begin{figure}[htbp]
    \centering
    \includegraphics[width=\linewidth]{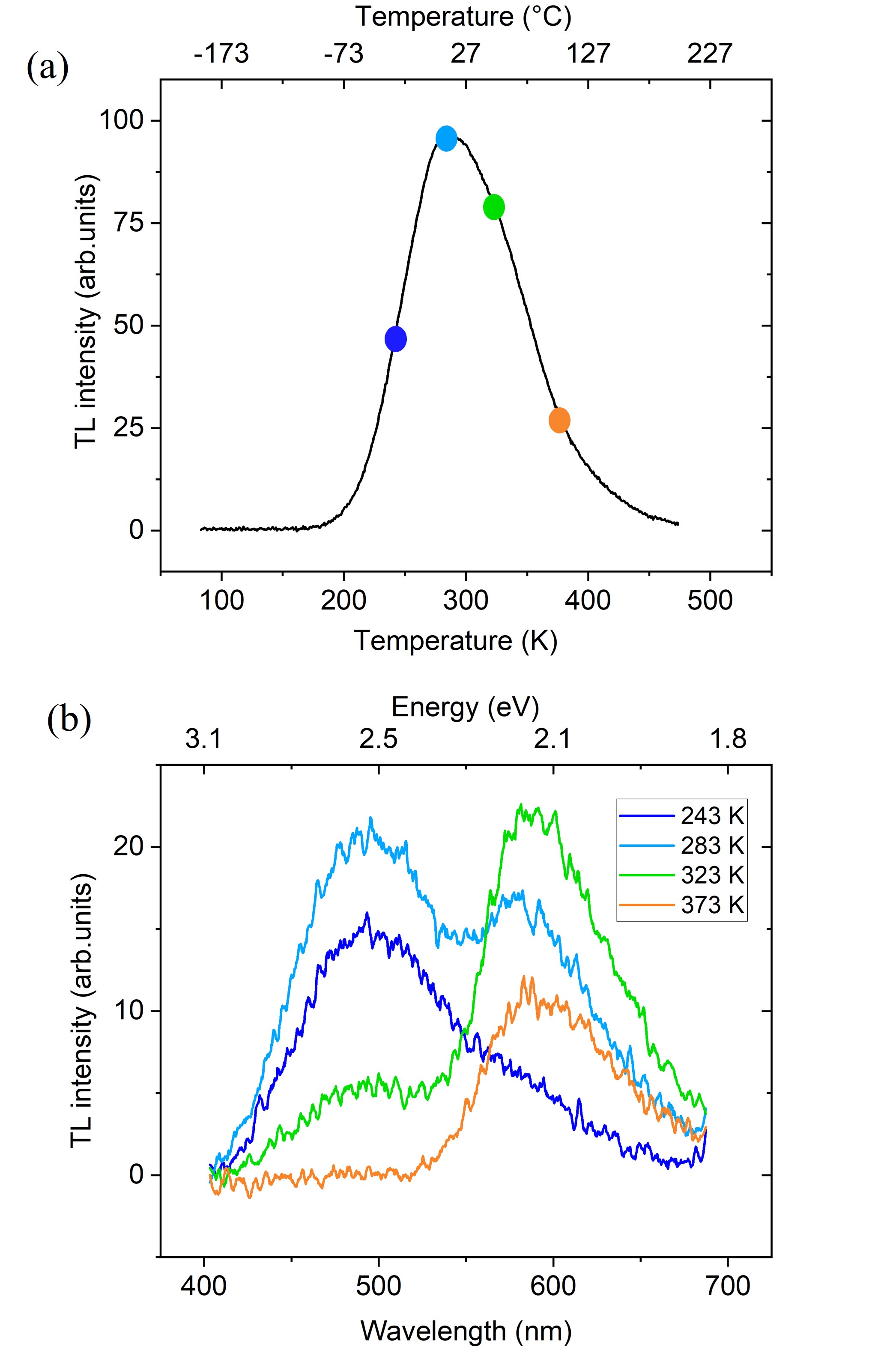}
    \caption{(a) TL glow curve recorded for sample Sino-01 after excitation with 224~nm laser at 83 K, delay for 5 min for phosphorescence to decay to zero, then heat to 473 K. Heating rate at 100 K/min. (b) Selected TL spectra: the line colours correspond to the colour of spots in (a) indicating the temperatures.}
    \label{fig:Sino-01 TL}
\end{figure}

TL glow curve for sample Sino-01 (Fig \textcolor{blue}{\ref{fig:Sino-01 TL}(a)}) starts to climb at around 173 K and reaches the maximum at approximately 283 K. The TL glow peak is asymmetric with a larger width on the high-temperature side indicating that it consists of more than one TL peaks \cite{mckeever1988thermoluminescence}. Two TL glow peaks are obtained by performing the TL cleaning process, which are centred at 284 K and 325 K, respectively. The activation energies of these two TL peaks calculated by initial rise method are 0.24(2) eV and 0.35(2)~eV.

A ``blue-green" band appears first at relatively low temperature (Fig \textcolor{blue}{\ref{fig:Sino-01 TL}(b)}), then as it fades, another ``orange" band appears. Its intensity gradually reaches about the same level as the maximum of the ``blue-green" band before finally disappearing when the temperature approaches 473 K. The ``blue-green" band is centred at 490 nm. There is no shift in the ``blue-green" band position during the thermoluminescence emission. By contrast, the centre of the ``orange" band shifts from approximately 580 nm to 590 nm as the temperature increases.

\subsubsection{Sample SYN4-10}

\begin{figure}[htbp]
    \centering
    \includegraphics[width=\linewidth]{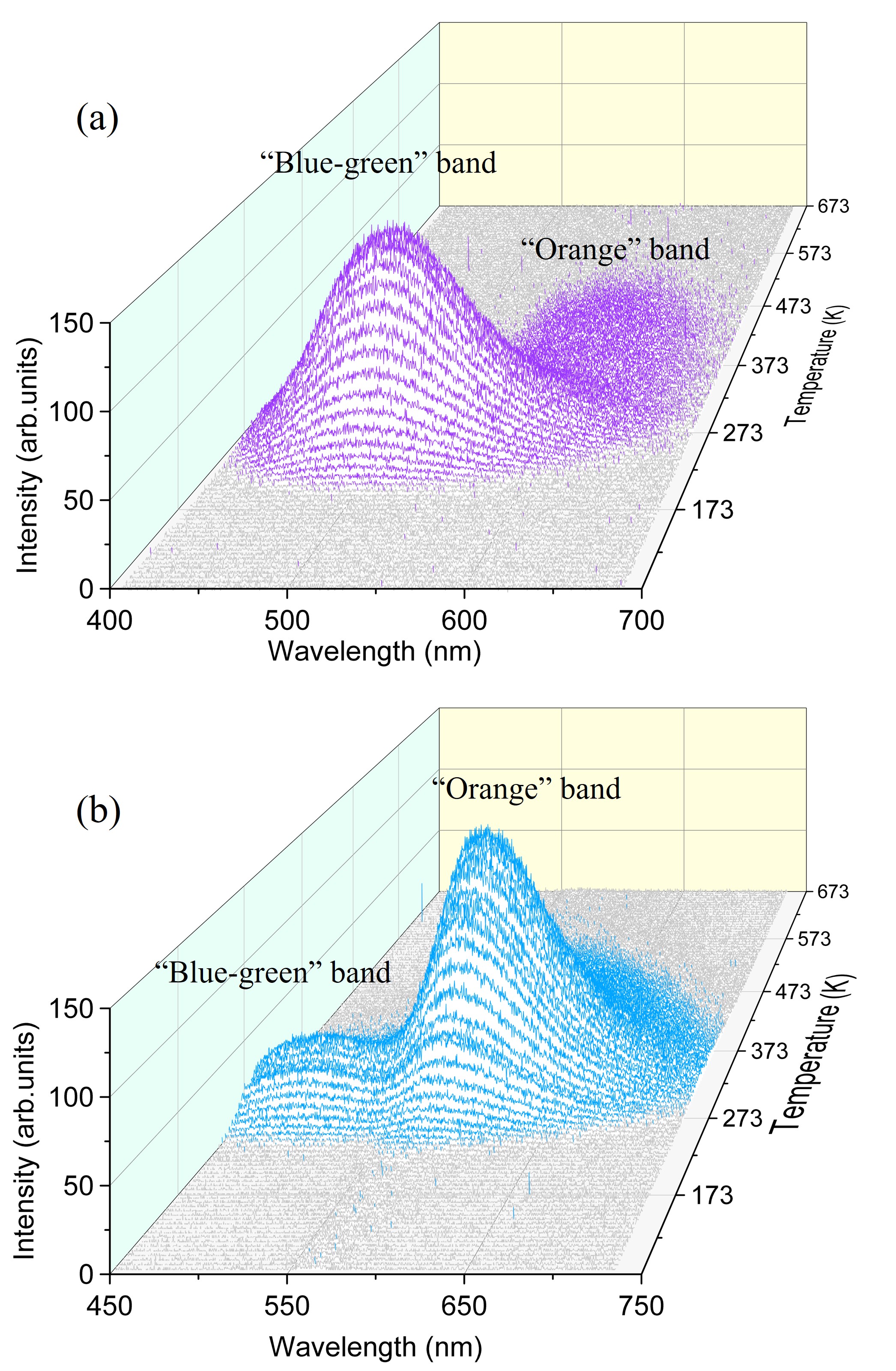}
      \caption{The 3D plot of thermoluminescence in SYN4-10 was recorded after excitation and phosphorescence from 83 K to 673 K at a 100 K/min heating rate. (a) after 224 nm pulsed laser excitation and 10 min delay (b) after 375 nm CW laser excitation and 10 min delay before ramping temperature.}
    \label{fig:SYN4-10 TL 3D plot}
\end{figure}

Fig \textcolor{blue}{\ref{fig:SYN4-10 TL 3D plot}} presents the thermoluminescence spectra in sample SYN4-10 following 224 nm and 375 nm laser excitation. In both cases, the thermoluminescence spectra include both a ``blue-green" and an ``orange" band. The ``blue-green" broadband appears first at lower temperatures, followed by an ``orange" band at higher temperatures. The relative intensities of the ``blue-green" and the ``orange" bands change dramatically when the wavelength of excitation changed from 224 nm to 375 nm.
 
The TL glow curves with both 224 and 375 nm excitation are broad and asymmetric suggesting multiple overlapping peaks. The initial rise method along with thermal cleaning results in activation energies ranging from 0.32(3) - 0.38(3) eV. The spectral position of the ``blue-green" TL band (centred at 490 nm) is temperature independent, but the ``orange" band shifts to longer wavelength (573 - 590 nm) with increasing temperature (Fig \textcolor{blue}{\ref{fig:SYN4-10 TL 3D plot}}).

The ``red" emission band observed in luminescence (Fig \textcolor{blue}{\ref{fig:Em vs Ex}}) and phosphorescence (Fig \textcolor{red}{\ref{fig:SYN4-10 375 phos 3Dplots}}) is not observed in the TL spectra of sample SYN4-10 following either 224 or 375 excitation (Fig \textcolor{blue}{\ref{fig:SYN4-10 TL 3D plot}}).

\subsection{\texorpdfstring{$\text{N}_\text{S}^0$} related charge transfer}

The $[{\text{N}_\text{S}^0}]$ was measured in two different growth sectors of the GE sample GE81-107a-B \{001\} and  GE81-107a-C \{111\} \cite{zhao2023}. In a \{111\} growth sector where the metastable sate (e.g. at room temperature after daylight/laboratory illumination for $>$ 1 hour), $[{\text{B}_\text{S}^0}] > [{\text{N}_\text{S}^0}]$, excitation with 224 nm light increased $[{\text{N}_\text{S}^0}]$, which subsequently decayed back to the metastable concentration after the excitation was removed. Whereas in a \{001\} growth sector where the metastable state $[{\text{N}_\text{S}^0}] > [{\text{B}_\text{S}^0}]$, excitation with 224 nm light decreased the $[{\text{N}_\text{S}^0}]$, which subsequently recovered after the excitation was removed \cite{zhao2023}. This data showed that for the substitutional nitrogen defect, three charge states have to be considered: negative, neutral, and positive \cite{zhao2023}.

\begin{figure}[htbp]
    \centering
    \includegraphics[width=\linewidth]{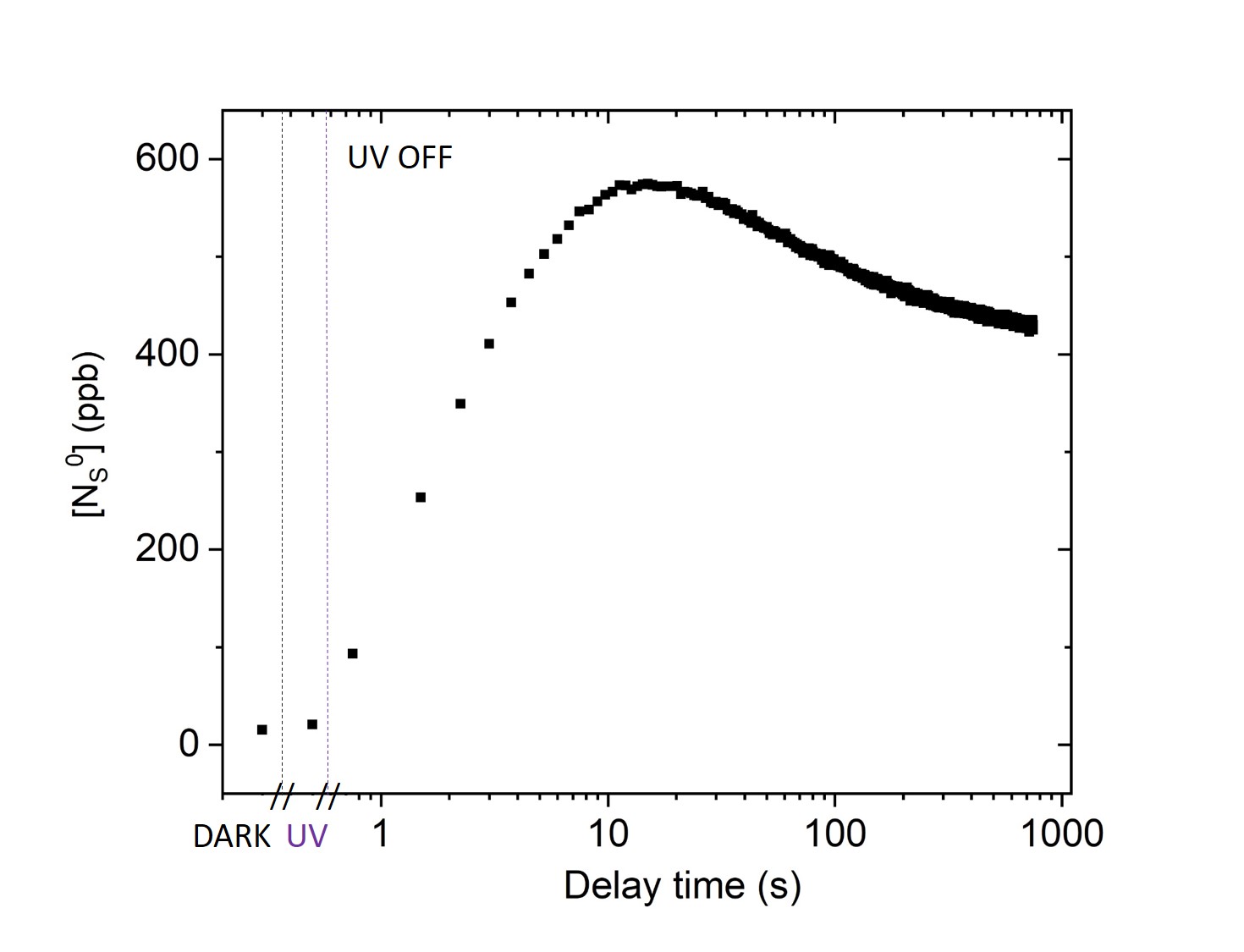}
    \caption{The ${\text{N}_\text{S}^0}$ concentration in sample SYN4-10 in dark, under UV excitation, and recovery-decay curve recorded after UV lamp excitation at room temperature.}
    \label{fig:SYN4-10 EPR}
\end{figure}

Above bandgap excitation increases the average concentration of ${\text{N}_\text{S}^0}$ in sample SYN4-10 from $15 \pm 2$ ppb to $21 \pm 2$ at room temperature when excited to ``saturation" ($[{\text{N}_\text{S}^0}]$ no longer changes). After turning off the UV lamp, the concentration of neutral substitutional nitrogen further increased for $\sim$ 10 s, then dropped at a slower rate (Fig. \textcolor{blue}{\ref{fig:SYN4-10 EPR}}). The decay of ${\text{N}_\text{S}^0}$ EPR signal lasts for $>$~12~hours.

\section{Discussion}

\subsection{Broad ``blue-green" luminescence and phosphorescence band mechanism}

It has been shown that the properties of the commonly observed broad ``blue-green" luminescence and phosphorescence band can be fully explained by emission from neutral substitutional nitrogen-boron donor-acceptor pairs ${\text{N}_\text{S}^0}$...${\text{B}_\text{S}^0}$, once the configurational change between charge states of the ${\text{N}_\text{S}}$ defect is considered, and both tunneling between defects and thermal ionization of donors and acceptors is included \cite{zhao2023}. ${\text{N}_\text{S}}$ and ${\text{B}_\text{S}}$ are the most abundant impurities in near colourless HPHT diamond, consistent with this assignment. Given the compact donor and acceptor wave functions in diamond, there are very few neighbours that should even be considered as ``close pairs". The vibrational broadening of the emission lines from very few close pairs means that they are undetectable, and the overlapping contributions from differently separated distant pairs (with different relaxation times) provide a natural explanation of the complex non-exponential decay. At low temperatures, when there is no thermal activation of carriers, the ${\text{N}_\text{S}^0}$...${\text{B}_\text{S}^0}$ phosphorescence will be dominated by relatively distant pairs, and the majority of pairs will emit only once. Thus the peak emission will be at slightly lower energies (2.25 eV). As the temperature increases, the probability of thermal ionization of isolated donors and acceptors increases and closer pairs can be reset by charge capture and emit multiple times. Furthermore, the long-lived distant pairs will be increasingly ionized before they can emit. Thus the emission peak will shift to higher energies (2.5 eV) as the temperature is increased. The magnitude of the shift observed is 0.25 eV, which is consistent with the Coulomb correction term \cite{zhao2023}. At low (tunneling of carriers dominates) and high (thermal excitation of carriers dominates) temperatures the peak of the broad ``blue-green" phosphorescence band does not shift with time, however in the intermediate temperature range (173 - 273 K), the phosphorescence band shifts to higher energy with time as expected.

The presence of both relatively shallow donors and acceptors is essential in order to reset  the nitrogen-boron donor-acceptor pairs into the ready-to-emit state ${\text{N}_\text{S}^0}$...${\text{B}_\text{S}^0}$ multiple times. This resetting of emitters facilitates bright and long-lived phosphorescence from a low concentration of relatively close donor-acceptor pairs. It must be remembered that in an insulating material like diamond, the calculated position of Fermi level does not necessarily predict the correct charge state of a defect, and the defect charge state is strongly influenced by the proximity of a particular defect to a donor (or acceptor). After or during above band gap excitation of near-colourless diamond doped with relatively low concentrations of substitutional nitrogen and boron defects, at temperatures where the probability of thermal ionization is low, there can be significant population of isolated ${\text{N}_\text{S}^-}$,  ${\text{N}_\text{S}^0}$ and  ${\text{B}_\text{S}^0}$ defects that persist almost indefinitely \cite{zhao2023}.

\subsection{Broad ``orange" luminescence and phosphorescence band mechanism}

In order to efficiently excite the ``blue-green" emission, excitation that produces free electrons and holes is required. At room temperature this threshold corresponds to an excitation wavelength shorter than 236 nm (Fig~\textcolor{blue}{\ref{fig:Em vs Ex}(a)}), excitation of an electron from the top of the valence band to the excitonic state
just below the conduction band, which can be thermally ionised. Of course lower energy photons could produce both free electrons and holes if there are mid-gap states that can accept an electron from the valence band (leaving a hole behind) and be ionized to liberate the electron to the conduction band. Such a process will be much less efficient than band gap excitation.

The broad featureless ``orange" luminescence and phosphorescence band (Fig \textcolor{blue}{\ref{fig:Em vs Ex}(b)}, \textcolor{blue}{\ref{fig:224 Phos spectra sino-01}, \textcolor{blue}{\ref{fig:Sino-01 TL}}}) centred on $\sim$ 580~nm (2.1 eV) at room temperature is produced by band gap excitation  but also by light of much longer wavelengths. At the same temperature, the lifetime of the ``orange" phosphorescence band is longer than that of the ``blue-green" phosphorescence. Although undetectable in sample Sino-01 at 223 K (Fig \textcolor{blue}{\ref{fig:224 Phos spectra sino-01}(a)}), the ``orange" phosphorescence band is stronger than the ``blue-green" band at 373 K (Fig \textcolor{blue}{\ref{fig:224 Phos spectra sino-01}(c)}). In sample SYN4-10, 224 nm excitation produces a strong ``blue-green" TL peak and a relatively weak ``orange" peak (Fig \textcolor{blue}{\ref{fig:SYN4-10 TL 3D plot}(a)}) but with 375 nm excitation the situation is reversed (Fig \textcolor{blue}{\ref{fig:SYN4-10 TL 3D plot}(b)}). Analysis of the glow curves in sample SYN4-10 following either 224 or 375 nm excitation gives effectively the same activation energies in the range 0.32 - 0.38 eV. Thus it could be inferred that one or even both of the traps involved in the ``blue-green" thermoluminescence/phosphorescence band are also involved in the emission from the ``orange" band.

The energy of the ``orange" thermoluminescence/phosphorescence band is not consistent with the emission from a ${\text{N}_\text{S}^-}$...${\text{B}_\text{S}^0}$ donor-acceptor pair which is expected to be in the UV range \cite{zhao2023}. The ``orange" thermoluminescence/phosphorescence band could arise from a different donor-acceptor pair e.g. ${\text{X}_\text{S}^-}$...${\text{B}_\text{S}^0}$ or ${\text{X}_\text{S}^0}$...${\text{B}_\text{S}^0}$, where X is another defect in the diamond. This appears unlikely as the unknown defect would need to be present in concentrations comparable to the nitrogen and/or boron impurities. 

Another possible explanation is a mechanism in which luminescence arises when a colour centre traps an electron or hole from the conduction or valence band. For example, consider the negatively charged (${\text{NV}^-}$) or neutral (${\text{NV}^0}$) nitrogen vacancy defects. For the NV defects there are three molecular orbitals (${\text{a}_1}$,${\text{e}_x}$,${\text{e}_y}$) in the band gap of diamond, the other ${\text{a}_1}$ state is pushed down into the valence band \cite{doherty2011}. The ${\text{NV}^-}$ defect's observable electronic structure consists of the ground ${\text{a}_1^2\text{e}^2}$ and first excited ${\text{a}_1^1\text{e}^3}$ molecular orbital configurations. Similarly, the molecular orbital configurations of ${\text{NV}^0}$ are ${\text{a}_1^2\text{e}^1}$ and ${\text{a}_1^1\text{e}^2}$ for the ground and first excited, respectively. Thus if ${\text{NV}^-}$ capture a hole from the valence band, it can arrive in an excited state of the neutral nitrogen vacancy defect 
\begin{equation}
{\text{NV}^-}:{\text{a}_1^2\text{e}^2} + {\text{h}^+_\text{VB}} \to {\text{NV}^0}:{\text{a}_1^1\text{e}^2}
\end{equation}
and subsequently emit a photon to transition to the ground sate of ${\text{NV}^0}$ \cite{sola2019,mizuochi2012}
\begin{equation}
{\text{a}_1^1\text{e}^2}  \to {\text{a}_1^2\text{e}^1} + \hbar\omega.
\end{equation}
${\text{NV}^0}$ can trap an electron from the conduction band but it must result in the ground state molecular orbital configuration of ${\text{NV}^-}$ defect
\begin{equation}
{\text{NV}^0}:{\text{a}_1^2\text{e}^1} + {\text{e}^-_\text{CB}} \to {\text{NV}^-}:{\text{a}_1^2\text{e}^2}
\end{equation}
and no emission is observed. This is why emission from ${\text{NV}^0}$ is observed in cathodoluminescence but the emission from ${\text{NV}^-}$ is not.

Imagine a mid gap defect X, that can exist in two charge states ${\text{X}^-}$ and ${\text{X}^0}$ and like the NV defect when ${\text{X}^-}$ traps a hole, an excited state of ${\text{X}^0}$ can be produced which relaxes to the ground state by ``orange" optical emission. Optical pumping of a diamond containing ${\text{N}_S}$, ${\text{B}_S}$, and X with 375 nm light could excite ``orange" photoluminescence from ${\text{X}^0}$ and it would excite electrons from ${\text{N}_S^0}$ to the conduction band. These electrons could be trapped by ${\text{X}^0}$ and produce ${\text{X}^-}$ and since ${\text{N}_S^0}$ also act as trap it would also be possible to produce populations of ${\text{N}_S^-}$, as well as ${\text{N}_S^+}$ and ${\text{X}^-}$. When the excitation is removed, at low temperatures we would expect a metastable population of ${\text{X}^-}$ to remain. As temperature increases, electrons are thermally excited to ${\text{B}_S^0}$, creating holes in the valence band which can be trapped by ${\text{X}^-}$ to produce ${\text{X}^0}$ in an excited state which thus give rise to ``orange" phosphorescence. Capture of an electron ionized from for example ${\text{N}_S^-}$, by ${\text{X}^0}$ would produce ${\text{X}^-}$ and the process could be repeated and long-lived phosphorescence observed. Bandgap optical excitation produces many free electrons and holes, and emission from ${\text{N}_\text{S}^0}$...${\text{B}_\text{S}^0}$ donor-acceptor pairs is observed as well as the ``orange" band emission. Optical pumping at wavelengths above 236 nm does not efficiently excite the ``blue-green" ${\text{N}_\text{S}^0}$...${\text{B}_\text{S}^0}$ donor-acceptor pair emission/phosphorescence and the different optical pumping efficiencies of 224 and 375 nm of ${\text{N}_\text{S}^0}$...${\text{B}_\text{S}^0}$ donor-acceptor pairs and X are clear from the change in intensity of the ``blue-green" and ``Orange" TL peaks (Fig \textcolor{blue}{\ref{fig:SYN4-10 TL 3D plot}}) and the phosphorescence spectra in figures~\textcolor{blue}{\ref{fig:224 Phos spectra syn4-10}} and \textcolor{blue}{\ref{fig:SYN4-10 375 phos 3Dplots}}. The fact that no zero phonon line is observed for band X emission indicates that there must be a large configurational change between the ground state and excited state of ${\text{X}^0}$ resulting a featureless broad emission band.

The observed luminescence, phosphorescence, and thermoluminescence data are consistent with invoking the existence of an X defect to explain the ``orange" emission band. It should be emphasised that this is not donor acceptor pair emission. This emission is an internal transition of an isolated defect, X need not be close to a donor or an acceptor, and the emission efficiency of ${\text{X}^0}$ could be very high such that only a very low concentration could give rise to long-lived phosphorescence. With the data available it is not possible to identify the constituents or structure of the postulated X defect. Given the presence of defects incorporating nickel in both Sino-01 and SYN4-10 but not the GE samples, it is tempting to suggest that X is a nickel related defect. Even for HPHT diamond samples grown in an iron-cobalt solvent-catalyst, trace contamination with nickel can result in the incorporation of a variety of nickel related defects. Since nickel defects are preferentially incorporated into \{111\} growth sectors it would be expected that the ``orange" emission should be found in \{111\} growth sectors which corresponding to what we observed in these two samples (see supplemental information). The increase in the ${\text{N}_S^0}$ concentration over $\sim$ 10 s, after bandgap excitation is removed in sample SYN4-10 (Fig \textcolor{blue}{\ref{fig:SYN4-10 EPR}}), is consistent with thermal ionization of ${\text{N}_\text{S}^-}$ and trapping by ${\text{X}^0}$ and ${\text{N}_\text{S}^+}$ defects.

\subsection{Broad ``red" luminescence and phosphorescence band mechanism} \label{sec:red}

\begin{figure}[htbp]
    \centering
    \includegraphics[width=\linewidth]{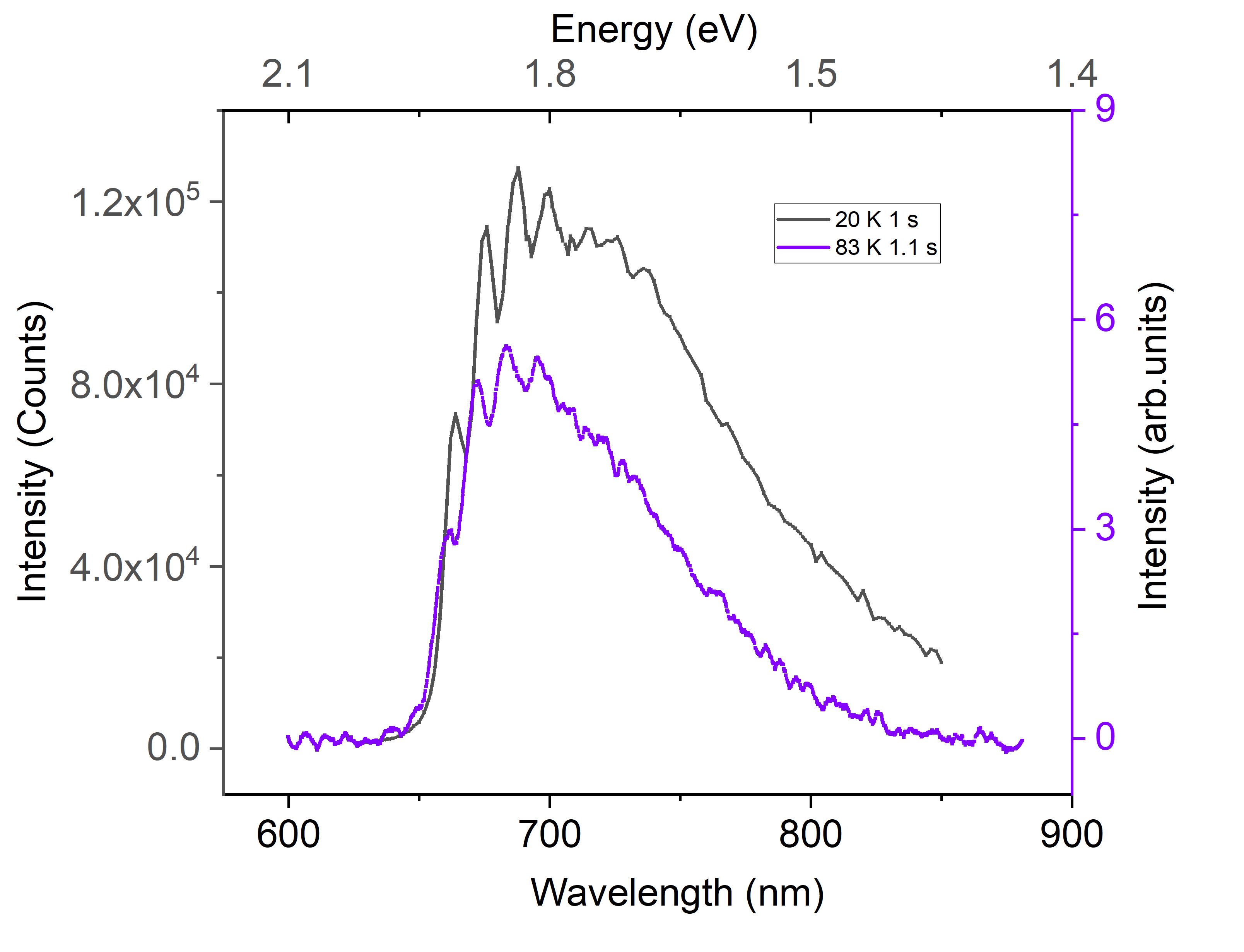}
    \caption{The phosphorescence spectra in sample SYN4-10 following 375 nm excitation recorded at 20 K and 83 K at a delay time of $\sim$ 1 s. At both temperatures, the vibronic replicas are with a phonon energy of $\sim$ 31 meV.}
    \label{fig:Red band T}
\end{figure}

An additional ``red" luminescent and phosphorescent band is observed in sample SYN4-10. This band is easily seen in the phosphorescence spectra excited at 375 nm rather than 224 nm. At room temperature the broad featureless ``red" phosphorescence band decays much more quickly than the ``orange" band (Fig \textcolor{blue}{\ref{fig:SYN4-10 375 phos 3Dplots}(b)}). At low temperatures, the ``red" emission band observed in SYN4-10 shows resolved vibronic replicas with a phonon energy of $\sim$ 31 meV (see Fig \textcolor{blue}{\ref{fig:Red band T}}). At 20 K the highest energy peak is at 1.868 eV (assigned to the zero phonon line) and this phosphorescence band persists for several seconds. Fitting the luminescence lineshape within the constraints of the 1D configuration coordinates model \cite{alkauskas2016} which uses a signal average phonon energy does not satisfactorily reproduce the shape of the emission band. Calculation of the full spectral function of the electron phonon coupling \cite{alkauskas2014} is beyond the scope of this current work but it is clear that in addition to the pronounced mode at $\sim$ 31 meV there must be a significant coupling to higher energy vibrational modes. For Ni containing defects, vibration frequency of order $\sim$ 31 meV have been previously recorded supporting the association with nickel \cite{yelisseyev2003}.

The resolved vibronic structures suggests that this emission originates from an isolated defect rather than donor-acceptor pairs. Interestingly, this red emission is not observed in the TL spectra, suggesting that the long-lived phosphorescence is not related to the generation and capture of free electrons and holes. It is possible that this is a phosphorescence system whereby after excitation and relaxation the electron is trapped in an excited energy level from which a transition to the ground state with the emission of a photon is forbidden (e.g. the levels have different spin quantum numbers). It has not been possible identify the defect responsible for the ``red" phosphorescence, but again nickel related defect is a possible candidate. Given that SYN4-10 is a brilliant cut multi-sector gemstone, it is impossible to determine which growth sectors the ``red" emission originates from. The lifetime of the red phosphorescence is possible growth sector dependent so it has not been measured. Further work would require single sector homogeneous samples to investigate the temperature dependence of the phosphorescence lifetime.

\section{Conclusions}

The physics of the ``blue-green", ``orange" and ``red" luminescence and phosphorescence bands in diamonds including the optical-excitation dependency (UV-NIR), temperature dependency (20 - 573 K), and related charge transfer process have been investigated, revealed that they have three different types of mechanisms. The ``blue-green"  luminescence and phosphorescence band can be fully explained by emission from neutral substitutional nitrogen-boron donor-acceptor pairs ${\text{N}_\text{S}^0}$...${\text{B}_\text{S}^0}$ recombination, and both tunneling between defects and thermal ionization of donors and acceptors is included. The long-lived ``orange" phosphorescence/thermoluminescence band can be explained by an alternative model based on capture of free carriers at an isolated colour centre X. Considering the strong distribution of the ``orange" luminescence and the  preferential incorporation of nickel related defects in the \{111\} growth sectors of diamond, nickel-related defects are temptingly suggested as candidates of X defects. For further work, optically detected magnetic resonance (ODMR) could be useful in the identification of X, if a long-lived excited state had non-zero electronic spin. The vibronic replicas with a phonon energy of $\sim$ 31 meV of ``red" phosphorescence band has been firstly observed at low temperatures in this paper, which suggests that this emission originates from an isolated defect when an electron is trapped in an excited energy level, the relaxation from this excited state to the ground state (emission of a photon) is forbidden.


\begin{acknowledgments}
We thank Prof Jon Goss of the Newcastle University for inspiring discussion. JZ thanks De Beers Ignite for providing funding. BLG gratefully acknowledges the Royal Academy of Engineering for a Research Fellowship. BLG and MEN acknowledge funding from EPSRC via grant EP/V056778/1.
\end{acknowledgments}

\nocite{*}

\bibliographystyle{apsrev4-2}
\bibliography{Delayed_luminescence_and_thermoluminescence_in_laboratory-grown_diamonds}

\end{document}